# A tutorial on conducting sample size and power calculations for detecting treatment effect heterogeneity in cluster randomized trials with linear mixed models

**Authors:** Mary Ryan Baumann[1,2*], Monica Taljaard[3,4], Patrick J. Heagerty[5], Michael O. Harhay[6], Guangyu Tong[7,8,9], Rui Wang[10,11], Fan Li[7,8,12]

[1]Department of Population Health Sciences, University of Wisconsin-Madison, Madison, WI, USA
[2]Department of Biostatistics and Medical Informatics, University of Wisconsin-Madison, Madison, WI, USA
[3]Methodological and Implementation Research, Ottawa Hospital Research Institute, Ottawa, ON, Canada
[4]School of Epidemiology and Public Health, University of Ottawa, Ottawa, ON, Canada
[5]Department of Biostatistics, School of Public Health, University of Washington, Seattle, WA, USA
[6]Center for Clinical Trials Innovation, Department of Biostatistics, Epidemiology and Informatics, Perelman School of Medicine, University of Pennsylvania, Philadelphia, PA, USA
[7]Section of Cardiovascular Medicine, Department of Internal Medicine, Yale School of Medicine, New Haven, CT, USA
[8]Department of Biostatistics, Yale School of Public Health, New Haven, CT, USA
[9]Cardiovascular Medicine Analytics Center, Yale School of Medicine, New Haven, CT, USA
[10]Department of Population Medicine, Harvard Pilgrim Health Care Institute, Boston, MA, USA
[11]Department of Biostatistics, Harvard T.H. Chan School of Public Health, Boston, MA
[12]Center for Methods in Implementation and Prevention Science, Yale School of Public Health, New Haven, CT, USA

***Corresponding author:**
Mary Ryan Baumann
Department of Population Health Sciences,
University of Wisconsin-Madison
WARF Office Building
610 Walnut Street
Madison, WI 53726
USA
mary.r.baumann@wisc.edu

## Abstract
Cluster-randomized trials (CRTs) are a well-established class of designs for evaluating community-based interventions. An essential task in planning these trials is determining the number of clusters and cluster sizes needed to achieve sufficient statistical power for detecting a clinically relevant effect size. While methods for evaluating the average treatment effect (ATE) for the entire study population are well-established, sample size methods for testing heterogeneity of treatment effects (HTEs), i.e., treatment-covariate interaction or difference in subpopulation-specific treatment effects, in CRTs have only recently been developed. For pre-specified analyses of HTEs in CRTs, effect-modifying covariates should, ideally, be accompanied by sample size or power calculations to ensure the trial has adequate power for the planned analyses. Power analysis for testing HTEs is more complex than for ATEs due to the additional design parameters that must be specified. Power and sample size formulas for testing HTEs via linear mixed effects (LME) models have been separately derived for different cluster-randomized designs, including single and multi-period parallel designs, crossover designs, and stepped-wedge designs, and for continuous and binary outcomes. This tutorial provides a consolidated reference guide for these methods and enhances their accessibility through an online R Shiny calculator. We further discuss key considerations for conducting sample size and power calculations to test pre-specified HTE hypotheses in CRTs, highlighting the importance of specifying advanced estimates of intracluster correlation coefficients for both outcomes and covariates, and their implications for power. The sample size methodology and calculator functionality are demonstrated through a real CRT example.

## Key messages

- Sample size and power calculations for studies investigating heterogeneity in treatment effects necessitate specifying intracluster correlation parameters for both outcomes and effect-modifying covariates to account for clustering.
- Adequately powering a cluster-randomized trial (CRT) for the overall treatment effect may simultaneously ensure sufficient sample size for testing pre-specified treatment effect heterogeneity for certain types of candidate effect-modifying covariates.
- Estimates for the required covariate and outcome intracluster correlation coefficients may be obtained from published reports or databases of completed CRTs or from analyses of available data from similar completed trials or observational studies.
- This paper provides a reference guide for the available sample size and power methods for different CRT designs assessing treatment effect heterogeneity and introduces an online calculator to facilitate practical application.

# Introduction

There are many types of cluster-randomized trials (CRTs) (Supplementary Figure S1), where intact groups of participants (clusters) are randomized to study conditions [1,2]; those under consideration in this tutorial, as well as associated terms, are defined in **Table 1**. While sample size calculation methods (i.e., number of clusters and cluster sizes) for detecting average treatment effects (ATEs) in CRTs have been thoroughly studied [3,4], guidance for assessing heterogeneity of treatment effects (HTEs) has only recently been developed for CRTs [5–12]. Here, we conceptualize an HTE as a specified treatment interaction with a continuous covariate or, in the case of a binary covariate, the difference in estimated treatment effects between two population subgroups.

*Table 1: Summary of types of cluster randomized trials (CRTs) under consideration, associated terms, and their definitions.*

| **Term** | **Definition** |
|---|---|
| Cluster | A naturally occurring group of individuals. |
| Subcluster | A cluster that is nested within a larger cluster. |
| CRT [13] | Cluster randomized trial – a trial where clusters are the unit of randomization instead of individuals, although the treatment itself may be administered to individuals or groups of individuals |
| Two-level parallel CRT [13] | Each cluster is randomized to one treatment condition for the duration of the trial such that intervention conditions have a concurrent control comparison (Supplementary Figure S1A). |
| Multiple-period parallel CRT | Clusters only experience one study condition but are measured repeatedly across several time periods. |
| Cluster randomized crossover (CRXO) trial [14] | Each cluster will be evaluated under both treatment conditions, with randomization determining the initial condition; all clusters synchronously switch to their next assigned treatment condition. |
| Two-period CRXO trial [14] | A CRXO trial with only one switch is known as a two-period CRXO trial, as each condition is considered for one (often equal-length) time period (Supplementary Figure S1B). |
| Multiple-period CRXO trial [15] | A CRXO trial where clusters can switch between conditions multiple times (Supplementary Figure S1C), or are observed for more than two time periods but only change treatment conditions once. |
| Stepped-wedge cluster randomized trial (SW-CRT) [16,17] | A CRXO trial where crossover direction is unidirectional (i.e., from control to treatment) and the crossover timing is randomized (Supplementary Figure S1D). |
| Three-level parallel CRT [18] | A parallel CRT where subclusters are nested within larger clusters. Randomization to treatment condition may occur at either the cluster or subcluster level. |
| Individual randomized group treatment (IRGT) trial [19] | A trial where individuals are the unit of randomization, but treatment conditions are administered in a clustered or grouped way (e.g., group instruction). |
| Step | The unique timepoints of a crossover in a CRXO trial. |
| Sequence | A unique treatment pattern a cluster may receive in a CRT. |
| Cross-sectional sampling [20] | Individuals are sampled from the population at a specific time period. If the CRT spans multiple periods, unique individuals are sampled from the population at each period. |
| Longitudinal/closed-cohort sampling [21] | Individuals are sampled from the population once (at baseline) and are repeatedly measured throughout the trial. |

There is growing interest in assessing the presence of HTE across equity-relevant subgroups and exploring intended or unintended differences in the effects of treatment, often even as a requirement for funding [22]. Although hypothesis-generating HTE analyses can be performed *post-hoc*, we focus on pre-specified analyses where effect-modifying covariates are identified *a priori*. In this context, understanding whether additional observations or participants are needed, beyond those required for assessing ATE, is essential. Power assessments for HTE analyses are often complex, as additional design parameters must be specified. Sample size formulas have been derived for several CRT designs – including single and multiple-period parallel designs with closed-cohort or cross-sectional sampling, crossover designs (CRXOs), and stepped-wedge designs (SW-CRTs) [5,11] – but without unification. This tutorial attempts to synthesize these methods and develop the CRT HTE Shiny Calculator (https://cluster-hte.shinyapps.io/shinyapp/).

To begin, we discuss sample size calculations in simple two-level CRTs to illustrate how methods for detecting HTEs are incorporated. Next, we introduce types of correlation structures – both for outcome and subgroup covariates – relevant to different designs and illustrate how more complex structures can impact sample size in a multiple-period CRT. We then demonstrate how our CRT HTE Shiny Calculator can be used to inform sample size calculations and explore implications of different assumptions for power. Finally, we provide recommendations for designing CRTs with HTE objectives and highlight potential areas of future research.

## Sample size calculation in parallel two-level CRTs

In a single-period parallel CRT with a continuous outcome, we can model HTE as an interaction between the cluster-level treatment variable $W_i$ and a potential effect-modifying covariate $X_{ik}$ via a linear mixed effects (LME) model:

$$Y_{ik} = \beta_0 + \beta_1 W_i + \beta_2 X_{ik} + \beta_3 W_i X_{ik} + \gamma_i + \epsilon_{ik},$$

where $Y_{ik}$ is the outcome for individual *k* in cluster *i* and $\epsilon_{ik} \sim N(0, \sigma_\epsilon^2)$ is the error. We define $X_{ik}$ as an individual-level variable; a cluster-level covariate can be used where $X_{ik} = X_i$. Outcome clustering is addressed via cluster-level random effects $\gamma_i \sim N(0, \sigma_\gamma^2)$. $\beta_1$ represents the treatment effect for individuals with $X_{ik} = 0$ (reference level; estimated by the generalized least squares estimator $\hat{\beta}_1$), while $\beta_1 + \beta_3 X_{ik}$ represents the treatment effect for individuals with $X_{ik} \neq 0$. Hence, the ATE can be defined as $\beta_1 + \beta_3 \mu_x$ where $\mu_x$ is the population mean of $X_{ik}$; in cases where $X_{ik}$ is mean-centered at 0, $\beta_1$ is the ATE. When assessing power to detect an HTE, we refer to whether the study has sufficient power to reject $\beta_3 = 0$ versus $\beta_3 = \Delta \neq 0$. For example, if $X_{ik} = 1$ when a participant has health insurance and 0 otherwise, $\beta_3$ represents the difference in treatment effects between the insured and uninsured subpopulations. Note that the magnitude of $\beta_3$ may be larger or smaller than $\beta_1$, depending on the magnitude and direction of the subpopulation-specific treatment effects.

The total sample size *N* is calculated as the number of clusters *n* multiplied by the number of individuals per cluster (i.e., cluster size) *m*. Thus, we can find the number of clusters such that:

$$n = \frac{\left(Z_{1-\alpha/2} + Z_{1-\beta}\right)^2 \sigma_*^2}{\Delta^2},$$

which relies on the significance level $\alpha$, effect size $\Delta$, power $1 - \beta$, and a variance $\sigma_*^2$ for a treatment effect estimator (ATE or HTE). Larger values of $\sigma_*^2$ will require a larger number of clusters.

If we target the ATE, $\sigma_*^2 = \sigma_{\text{ATE}}^2$. In parallel two-level CRT settings this is given by [5,7]:

$$\sigma_{\text{ATE}}^2 = \frac{\sigma_{y|x}^2 \{1 + (m-1)\alpha_1\}}{m\pi(1-\pi)\sigma_x^2},$$

which depends on *m*, the conditional outcome variance $\sigma_{y|x}^2$, variance of the covariate $\sigma_x^2$ (or $(\text{prevalence of } x) \times (1 - \text{prevalence of } x)$ for binary $X_{ik}$), the proportion of clusters randomized to treatment $\pi$, and $\alpha_1$, a measure of outcome clustering known as the outcome intracluster correlation (ICC).

Further, if we target treatment effect modification, $\sigma_*^2 = \sigma_{\text{HTE}}^2$:

$$\sigma^2_{\text{HTE}} = \sigma^2_{\text{ATE}} \times \underbrace{\frac{(1-\alpha_1)}{\{1+(m-2)\alpha_1-(m-1)\rho_1\alpha_1\}}}_{\text{HTE design effect}}.$$

Thus, the "design effect" of a study focused on HTE hypotheses depends on both the outcome ICC $\alpha_1$ and *covariate* ICC $\rho_1$. **Table 2** outlines a summary of available HTE variance formulas and design effects for various study designs.

*Table 2: Summary of heterogeneity of treatment effect (HTE) variance formulas for cluster randomized trials (CRTs) by design type, number of time periods, and sampling scheme. Abbreviations: cluster randomized crossover (CRXO); intracluster correlation (ICC); individually-randomized group treatment (IRGT); stepped-wedge CRT (SW-CRT).*

| **Design** | **Time periods** ($J$) | **Sampling Scheme** | **Variance expression** |
|---|---|---|---|
| Parallel | One | Two-level [5] | $\frac{\sigma_{y\mid x}^2}{\pi(1-\pi)} \times \frac{(1-\alpha_1)\{1+(m-2)\alpha_1\}}{m\{1+(m-2)\alpha_1-(m-1)\rho_1\alpha_1\}}$ |
| | One | Three-level [6] | Cluster-level randomization: $\frac{\sigma_{y\mid x}^2}{\pi(1-\pi)\sigma_x^2} \times \frac{n_s m}{\lambda_3(\alpha_0,\alpha_1)^{-1}\zeta_3(\rho_0,\rho_1)+(n_s-1)\lambda_2(\alpha_0,\alpha_1)^{-1}\zeta_2(\rho_0,\rho_1)+n_s(m-1)\lambda_1(\alpha_0)^{-1}\zeta_1(\rho_0)}$<br>Subcluster-level randomization: $\frac{\sigma_{y\mid x}^2}{\pi(1-\pi)\sigma_x^2} \times \frac{m}{m\lambda_1(\alpha_0)^{-1}-\{1+(m-1)\rho_0\}(\lambda_1(\alpha_0)^{-1}-\lambda_2(\alpha_0,\alpha_1)^{-1})}$ |
| CRXO | Multiple | Cross-sectional [12] | $\frac{\sigma_{y\mid x}^2}{\pi(1-\pi)\sigma_x^2} \times \left\{\frac{2(J-1)\zeta_1(\rho_0)}{\lambda_1(\alpha_0)} + \frac{\zeta_3(\rho_0,\rho_1)}{\lambda_2(\alpha_0,\alpha_1)} + \frac{\zeta_2(\rho_0,\rho_1)}{\lambda_3(\alpha_0,\alpha_1)}\right\}^{-1}$ |
| | Multiple | Cohort [12] | $\frac{\sigma_{y\mid x}^2}{\pi(1-\pi)\sigma_x^2} \times \left[2\left\{\frac{(J-1)\eta_1(\rho_0)}{\tau_1(\alpha_0,\alpha_1,\alpha_2)} + \frac{\eta_2(\rho_0)}{\tau_3(\alpha_0,\alpha_1,\alpha_2)}\right\}\right]^{-1}$ |
| SW-CRT | Multiple | Cross-sectional [11] | $\frac{\sigma_{y\mid x}^2/\sigma_x^2}{tr(\mathbf{\Omega}_W)} \times \frac{J^2}{n(J-1)(1-\tau_W)(\zeta_3(\rho_0,\rho_1)-\zeta_2(\rho_0,\rho_1))(\lambda_2(\alpha_0,\alpha_1)^{-1}-\lambda_3(\alpha_0,\alpha_1)^{-1})+J\theta^{CS}(J,m)}$ |
| | Multiple | Cohort [11] | $\frac{\sigma_{y\mid x}^2/\sigma_x^2}{tr(\mathbf{\Omega}_W)}$<br>$\times \frac{J^2}{n(J-1)(1-\tau_W)\{(\tau_3(\alpha_0,\alpha_1,\alpha_2)^{-1}-\tau_4(\alpha_0,\alpha_1,\alpha_2)^{-1})\eta_2(\rho_0)+(N-1)(\tau_1(\alpha_0,\alpha_1,\alpha_2)^{-1}-\tau_2(\alpha_0,\alpha_1,\alpha_2)^{-1})\eta_1(\rho_0)\}+J\theta^{CC}(J,m)}$ |
| IRGT | One | Two-level [8] | Individual-level covariate: $\frac{\sigma_1^2(1-\alpha_{1,1})\{1+(m_1-1)\alpha_{1,1}\}}{\sigma_x^2\pi m_1\{1+(m_1-2)\alpha_{1,1}\}} + \frac{\sigma_0^2(1-\alpha_{1,0})\{1+(m_0-1)\alpha_{1,0}\}}{\sigma_x^2(1-\pi)m_0\{1+(m_0-2)\alpha_{1,0}\}}$<br>Cluster-level covariate: $\frac{\sigma_1^2\{1+(m_1-1)\alpha_{1,1}\}}{\sigma_x^2\pi m_1} + \frac{\sigma_0^2\{1+(m_0-1)\alpha_{1,0}\}}{\sigma_x^2(1-\pi)m_0}$ |

*$n$: number of clusters; $m$: cluster size; $J$: number of periods; $\alpha_*$: outcome ICC; $\alpha_{*,\#}$: outcome ICC for treatment group #; $\rho_*$: covariate ICC; $\rho_{*,\#}$: covariate ICC for treatment group #; $\pi$: proportion of clusters to be randomized to treatment group/sequence; $\lambda_*(\cdot)$: eigenvalues of the nested exchangeable outcome correlation structure; $\zeta_*(\cdot)$: eigenvalues of the nested exchangeable covariate correlation structure; $\tau_W$: generalized ICC of the intervention vector; $\tau_*(\cdot)$: eigenvalues of the block exchangeable outcome correlation structure; $\eta_*(\cdot)$: eigenvalues of the exchangeable covariate correlation structure; $tr(\mathbf{\Omega}_W)$: trace of the covariance matrix of the intervention vector; $\theta^{CS}(\cdot)$: the largest eigenvalue of a nested exchangeable correlation matrix; $\theta^{CC}(\cdot)$: the largest eigenvalue of an exchangeable correlation matrix. For mathematical definitions of terms, see references provided in the "Sampling Scheme" column.*

## Intracluster correlation structures

Sample size calculations for CRTs require information about how strongly outcomes are correlated within clusters as measured through the outcome ICC: the ratio of between-cluster and total outcome variation. As the data clustering becomes more complex, several outcome ICC structures can be specified, including exchangeable (common ICC $\alpha_1 = corr(Y_{ik}, Y_{ik'})$), nested exchangeable (within-period ICC $\alpha_1 = corr(Y_{ijk}, Y_{ijk'})$, between-period ICC $\alpha_2 = corr(Y_{ijk}, Y_{ij'k'})$), and block exchangeable (including an additional within-individual ICC $\alpha_0 = corr(Y_{ijk}, Y_{ij'k})$ to describe associations between repeated measures). As in-depth overviews of these structures are elsewhere in the literature [4,23,24], we summarize classifications in **Table 3** with illustrations in the Supplementary Material (Supplementary Figure S2). The between-period ICC $\alpha_2$ can also be parameterized via a cluster autocorrelation coefficient (CAC): the ratio of between-period and within-period ICCs ($\alpha_2/\alpha_1$) [25]. Generally, lower-level relationships in the structure (e.g., within-individual ICCs) are smaller than or equal to higher-level relationships (e.g., between-period ICCs).

*Table 3: Taxonomy of available outcome correlation structures for treatment effect heterogeneity-focused cluster randomized trials (CRTs) by number of time periods, sampling scheme, levels of clustering, and types of correlation. Abbreviations: correlation (corr); intracluster correlation (ICC).*

| **Time periods** | **Sampling scheme** | **Number of clustering levels** | **Maximum outcome correlation structure complexity** | **Correlation components** | **Outcome model** |
|---|---|---|---|---|---|
| Single | Parallel | Two | Exchangeable | ICC | $\alpha_1 = corr(Y_{ik}, Y_{ik'})$ |
| | | Differential clustering by arm | Arm-specific exchangeable | ICC (control)<br>ICC (treatment) | $\alpha_1^{\text{control}} = corr(Y_{ik}^{\text{control}}, Y_{ik'}^{\text{control}})$<br>$\alpha_1^{\text{trt}} = corr(Y_{ik}^{\text{trt}}, Y_{ik'}^{\text{trt}})$ |
| | | Three | Nested exchangeable | Within-cluster<br>Between-cluster | $\alpha_1 = corr(Y_{ijk}, Y_{ijk'})$<br>$\alpha_2 = corr(Y_{ijk}, Y_{ij'k'})$ |
| Multiple | Cross-sectional | Two | Nested exchangeable | Within-period<br>Between-period | $\alpha_1 = corr(Y_{ijk}, Y_{ijk'})$<br>$\alpha_2 = corr(Y_{ijk}, Y_{ij'k'})$ |
| | Closed-cohort/ longitudinal | Three | Block exchangeable | Within-individual<br>Within-period<br>Between-period | $\alpha_0 = corr(Y_{ijk}, Y_{ij'k})$<br>$\alpha_1 = corr(Y_{ijk}, Y_{ijk'})$<br>$\alpha_2 = corr(Y_{ijk}, Y_{ij'k'})$ |

While a correlation structure for the outcome is required regardless of whether the sample size calculation is performed for the ATE or HTE, an additional structure is necessary for the HTE due to potential clustering of *covariates* [5,26]. Covariate correlation, represented by $\rho$, can take on similar structures to outcomes, depending on the design type. The covariate ICC for an individual-level covariate typically carries similar magnitude to the outcome ICC, whereas $\rho = 1$ for a cluster-level covariate. Similar to the outcome ICC structure, one needs to differentiate between the within-period ($\rho_1$) and between-period ICCs ($\rho_2$) in multiple-period CRTs. If the covariate is a cluster-level characteristic (e.g., rurality), the within-period and between-period covariate ICCs are $\rho_1 = corr\left(X_{ijk}, X_{ijk'}\right) = 1$ and $\rho_2 = corr\left(X_{ijk}, X_{ij'k'}\right) = 1$, respectively, even though the outcome ICC structure is nested exchangeable ($\alpha_1 \neq \alpha_2$). Further discussion of appropriate nested exchangeable covariate correlation structures for three-level and stepped-wedge designs can be found in Li et al. (2022) and (2024) [6,11]. The correspondence between design types and appropriate outcome and covariate correlation structures is summarized in **Table 4**.

*Table 4: Alignment of maximally complex outcome and covariate correlation structures by unit of randomization, design type, and sampling scheme. Abbreviations: cluster randomized trial (CRT); cluster randomized crossover trial (CRXO).*

| **Unit of randomization** | **Design** | **Sampling scheme** | **Outcome correlation structure** | **Covariate correlation structure** |
|---|---|---|---|---|
| Cluster | Two-level CRT | Parallel | Exchangeable | Exchangeable |
| | Two-level CRT with differential clustering by arm | | Arm-specific exchangeable | Exchangeable |
| | Multiple-period CRT (parallel or CRXO) | Cross-sectional | Nested exchangeable | Nested exchangeable |
| | | Closed-cohort/ longitudinal | Block exchangeable | Exchangeable |
| Cluster/ subcluster | Three-level CRT | Parallel | Nested exchangeable | Nested exchangeable |
| Individual | Individually-randomized group treatment | | Arm-specific exchangeable | Independent |

In ATE-focused studies, not allowing for distinct between-period ICCs in multiple-period CRTs results in artificially small sample sizes [27]; this makes selection of appropriately flexible correlation structures crucial in sample size estimation. We next discuss HTE sample size calculations for multiple-period CRTs to demonstrate how more complex structures are incorporated.

## Sample size calculations in multiple-period CRTs

Multiple-period CRTs can use cross-sectional or longitudinal closed-cohort sampling to measure individuals within clusters. HTEs in parallel multiple-period cross-sectional CRTs may be modeled as interactions between treatment $W_{ij}$ for cluster *i* in period *j* and covariate $X_{ijk}$ for individual *k* in cluster *i* at period *j* via an LME:

$$Y_{ijk} = \beta_{0j} + \beta_1 W_{ij} + \beta_{2j} X_{ijk} + \beta_3 W_{ij} X_{ijk} + \gamma_i + \eta_{ij} + \epsilon_{ijk}.$$

Note that now we have period-specific intercept terms ($\beta_{0j}$) that account for time trends, period-specific covariate terms ($\beta_{2j}$), and cluster-by-period random effects $\eta_{ij} \sim N(0, \sigma_\eta^2)$ [12]. Multiple-period cross-sectional CRTs include both within-period and between-period comparisons, making nested exchangeable correlation structures necessary for both the outcome and covariate. The variances of ATE and HTE estimators for a multiple-period CRT are complex, involving terms for the number of periods *J*, cluster-period size and number of clusters (*m, n*), outcome variance $\sigma_{y|x}^2$, covariate variance $\sigma_x^2$, and within-period and between-period outcome ICCs ($\alpha_1, \alpha_2$). For CRXOs, the proportion of clusters randomized to the treatment sequence starting with treatment in the first period ($\pi$), is also needed. The HTE variance (**Table 2**) additionally involves terms for the within- and between-period covariate ICCs ($\rho_1, \rho_2$) [11,12].

To accommodate closed-cohort longitudinal sampling of participants, additional cluster-by-individual random effects $s_{ik} \sim N(0, \sigma_s^2)$ are added to the above model:

$$Y_{ijk} = \beta_{0j} + \beta_1 W_{ij} + \beta_{2j} X_{ijk} + \beta_3 W_{ij} X_{ijk} + \gamma_i + \eta_{ij} + s_{ik} + \epsilon_{ijk}.$$

In addition to the terms needed for the cross-sectional setting, ATE and HTE variances also require within-individual outcome and covariate ICCs ($\alpha_0, \rho_0$; block exchangeable structure) to address within-individual comparisons (see **Table 2**). We discuss implications of the HTE variance components on sample size in the Supplementary Material.

## Sample size calculation workflow and the CRT HTE Shiny Calculator

We implemented sample size calculation methods for the previously discussed designs in a free online application, the CRT HTE Shiny Calculator: https://cluster-hte.shinyapps.io/shinyapp/. Designs supported by the calculator include: parallel two-level, three-level, and multiple-period CRTs; two-period and multiple-period CRXOs; SW-CRTs; individually randomized group treatment trials (IRGTs); and parallel two-level CRTs with by-arm heterogeneity in sample size and ICC specifications. Users may also upload a design via a CSV file. Parallel designs require a randomization proportion of clusters to treatment conditions. Crossover or multiple-period designs require specification of how individuals within a cluster will be sampled across time (cross-sectional versus closed-cohort) as well as the number of periods, while SW-CRTs also require the number of sequences, assuming clusters are balanced across sequences.

Investigators must also provide information about the outcome and covariate, including data type (continuous, binary) and correlation estimates (outcome $\alpha_1$, CAC, $\alpha_0$ if closed-cohort; covariate $\rho_1$, CAC). The calculator uses a linear approximation for binary outcomes, targeting risk differences [9]. Data type will determine whether investigators must provide standard deviations (outcome: $\sigma_{y|x}$; covariate: $\sigma_x$) for continuous variables, or prevalences for binary variables. A postulated or minimum clinically meaningful HTE size is also required.

The calculator determines the most complex correlation structure available based on the chosen design and sampling scheme and prompts for the appropriate information. Users may also conduct sensitivity analyses by inputting minimum and maximum ICCs.

Finally, investigators must provide two of three quantities: number of clusters, cluster-period size, or power. The calculator outputs plots of sample size and power curves, such that one parameter is fixed, a range is specified for the second, and the third is solved for. Results can be viewed in four ways: cluster size versus power (number of clusters fixed), number of clusters versus power (cluster-period size fixed), cluster size versus number of clusters (power fixed), or HTE size versus power (number of clusters and cluster-size fixed).

Next, we demonstrate the application of the calculator to a real-world CRT. An additional example and discussion of other practical considerations (obtaining initial estimates of ICCs, incorporating small sample corrections and variable cluster sizes, obtaining power for an HTE when sample size is based on detecting the ATE), can be found in the Supplementary Material.

## Data example: exploration of a SW-CRT via the Lumbar Imaging with Reporting of Epidemiology (LIRE) study

LIRE is a SW-CRT in clinics that tested the effect of adding prevalence data for common imaging findings in patients without back pain to lumbar spine imaging reports received by primary care [28]. The primary outcome (continuous) was spine-related intervention intensity based on Relative Value Units during the year following imaging.

Imaging may be performed using either plain film or advanced imaging techniques – suppose a trial is being planned to estimate the difference in treatment effects between reports where plain film imaging is used and advanced imaging reports. The planned design is a cross-sectional SW-CRT, using nested exchangeable structures for both outcome and covariate correlations. To estimate the required number of patients per clinic-period (*m*), the parameters needed are: number of clinics (*n*), number of treatment sequences (*J-1*), within-period outcome ICC ($\alpha_1$), outcome CAC ($\alpha_2/\alpha_1$), within-period covariate ICC ($\rho_1$), and covariate CAC ($\rho_2/\rho_1$), covariate prevalence, power, and HTE size.

The original LIRE trial planned to recruit 100 clinics across 5 sequences (6 six-month periods), with an average of 418 patients per clinic-period ($N = 250{,}800$). We assume a similar study length and number of clusters for our study. A previously published report provides an overall $\alpha = 0.013$ with 95% confidence interval (0,0.046) [28]. ICCs obtained from exchangeable structures are often smaller than those in nested exchangeable structures, and there is little relevant information to inform our outcome CAC; therefore, we used a Shiny web application [29] to calculate $\alpha_1$ and outcome CAC value pairs consistent with the above $\alpha$ range: $(\alpha_1, CAC) = (0.022, 0.5)$, with the lower and upper ends of the range defined by $(0.014, 0.9)$ and $(0.046, 0.13)$. Using LIRE study data, we estimate 20% prevalence of advanced imaging. Lacking specific prior information, we conservatively use $\rho_1 = 0.1$ (range: 0.05-0.15) and $\rho_2/\rho_1 = 0.9$ (range: 0.7-1) as the ICC for binary covariates is constrained by the prevalence [30,31]. A reasonable target standardized HTE size is -0.05 (dividing by the outcome standard deviation), which represents 50% of the assumed standardized ATE size of -0.1 . Finally, we assumed a two-sided 5% type I error.

Under these assumptions, 100 clinics – each with a cluster-period size of 353 patients ($n \times m \times J = N = 211{,}800$) – would provide 90% power to detect the anticipated HTE size (**Figure 1**). It may also be of interest to explore a multiple-period parallel CRT to better understand trade-offs of resources, logistics, and potential bias [32]. If we were to use a balanced six-period parallel CRT, a much smaller cluster-period size of 190 ($n = 100$ total clinics under 1:1 allocation; $N = 114{,}000$) would be required to reach 90% power for the HTE, while a similar CRXO design would require a cluster-period size of 185 ($N = 111{,}000$); both exhibit similar scales of sample size reduction one would see if designing similarly-parameterized trials for the ATE.

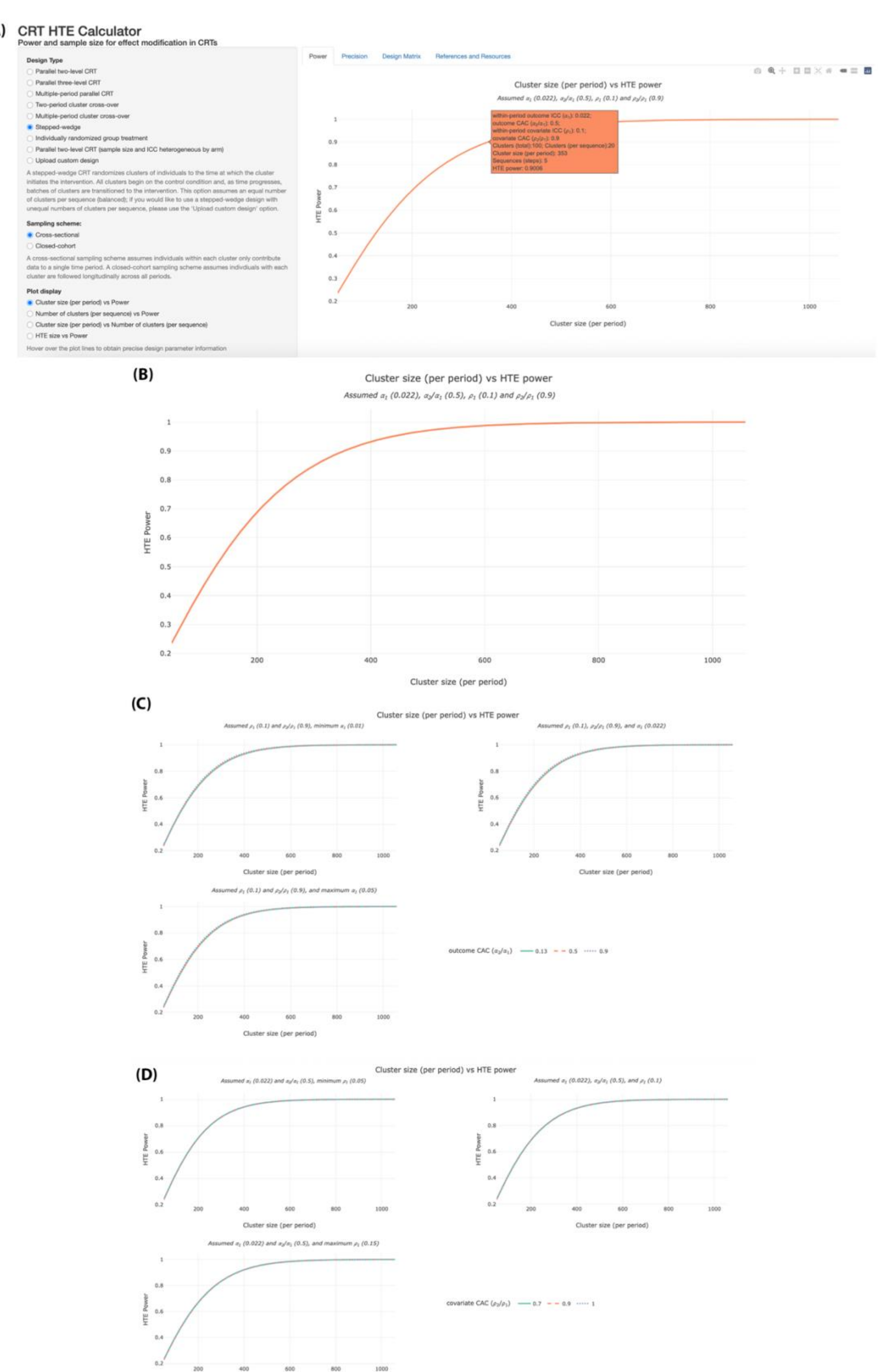
(A)
CRT HTE Calculator
Power and sample size for effect modification in CRTs
Design Type
Parallel two-level CRT
Parallel three-level CRT
Multiple-period parallel CRT
Two-period cluster cross-over
Multiple-period cluster cross-over
Stepped-wedge
Individually randomized group treatment
Parallel two-level CRT (sample size and ICC heterogeneous by arm)
Upload custom design
A stepped-wedge CRT randomizes clusters of individuals to the time at which the cluster initiates the intervention. All clusters begin on the control condition and, as time progresses, batches of clusters are transitioned to the intervention. This option assumes an equal number of clusters per sequence (balanced); if you would like to use a stepped-wedge design with unequal numbers of clusters per sequence, please use the 'Upload custom design' option.
Sampling scheme:
Cross-sectional
Closed-cohort
A cross-sectional sampling scheme assumes individuals within each cluster only contribute data to a single time period. A closed-cohort sampling scheme assumes indivdiuals with each cluster are followed longitudinally across all periods.
Plot display
Cluster size (per period) vs Power
Number of clusters (per sequence) vs Power
Cluster size (per period) vs Number of clusters (per sequence)
HTE size vs Power
Hover over the plot lines to obtain precise design parameter information
Power
Precision
Design Matrix
References and Resources
Cluster size (per period) vs HTE power
Assumed $\alpha_1$ (0.022), $\alpha_2/\alpha_1$ (0.5), $\rho_1$ (0.1) and $\rho_2/\rho_1$ (0.9)
within-period outcome ICC ($\alpha_1$): 0.022;
outcome CAC ($\alpha_2/\alpha_1$): 0.5;
within-period covariate ICC ($\rho_1$): 0.1;
covariate CAC ($\rho_2/\rho_1$): 0.9
Clusters (total):100; Clusters (per sequence):20
Cluster size (per period): 353
Sequences (steps): 5
HTE power: 0.9006
HTE Power
Cluster size (per period)
(B)
Cluster size (per period) vs HTE power
Assumed $\alpha_1$ (0.022), $\alpha_2/\alpha_1$ (0.5), $\rho_1$ (0.1) and $\rho_2/\rho_1$ (0.9)
HTE Power
Cluster size (per period)
(C)
Cluster size (per period) vs HTE power
Assumed $\rho_1$ (0.1) and $\rho_2/\rho_1$ (0.9), minimum $\alpha_1$ (0.01)
Assumed $\rho_1$ (0.1), $\rho_2/\rho_1$ (0.9), and $\alpha_1$ (0.022)
Assumed $\rho_1$ (0.1) and $\rho_2/\rho_1$ (0.9), and maximum $\alpha_1$ (0.05)
HTE Power
Cluster size (per period)
outcome CAC ($\alpha_2/\alpha_1$) 0.13 0.5 0.9
(D)
Cluster size (per period) vs HTE power
Assumed $\alpha_1$ (0.022) and $\alpha_2/\alpha_1$ (0.5), minimum $\rho_1$ (0.05)
Assumed $\alpha_1$ (0.022), $\alpha_2/\alpha_1$ (0.5), and $\rho_1$ (0.1)
Assumed $\alpha_1$ (0.022) and $\alpha_2/\alpha_1$ (0.5), and maximum $\rho_1$ (0.15)
HTE Power
Cluster size (per period)
covariate CAC ($\rho_2/\rho_1$) 0.7 0.9 1

*Figure 1: Screenshot of the CRT HTE Shiny Calculator interface and power curves from calculator for a stepped-wedge cluster randomized trial (SW-CRT) with a continuous outcome and a binary effect modify covariate, modeled after the Lumbar Imaging with Reporting of Epidemiology (LIRE) study. Scenario includes 5 sequences, 20 clusters per sequence, outcome Intracluster correlation (ICC) of 0.022 (lower value 0.01 and higher value 0.05), outcome cluster autocorrelation (CAC) of 0.5 (lower value 0.13 and higher value 0.9), covariate ICC of 0.1 (lower value 0.05 and higher value 0.15), covariate CAC of 0.9 (lower value 0.7 and higher value 1), standardized heterogeneity treatment effect (HTE) size of -0.05, effect modifier prevalence of 20%, and 0.05 significance level. Figure 1A shows a screenshot of a portion of the Calculator interface. Figure 1B depicts curve assuming fixed within-period outcome ICC of 0.022, outcome CAC of 0.5, within-period covariate ICC of 0.1, and covariate CAC of 0.9. Figure 1C depicts multiple curves assuming within-period outcome ICC and CAC ranges of (0.01, 0.05) and (0.13, 0.9), respectively, a fixed within-period covariate ICC of 0.5, and a fixed covariate CAC of 0.9. Figure 1D depicts multiple curves assuming within-period covariate ICC and CAC ranges of (0.05, 0.15) and (0.7, 1), respectively, a fixed within-period outcome ICC of 0.022, and a fixed outcome CAC of 0.5.*

# Discussion

In this tutorial, we considered how assumed estimates for outcome and covariate clustering can be used to plan CRT designs that study heterogeneity of treatment effects. We presented our CRT HTE Shiny Calculator, which allows investigators to easily obtain sample size and power estimates for CRTs with pre-planned effect modification hypotheses using LMEs. In any CRT investigating HTEs, it is crucial to account for clustering in both the subgroup covariate as well as the outcome. However, obtaining estimates of the covariate ICC may be simpler than the outcome ICC as some covariates may appear more commonly in a wider range of studies. In the Supplementary Material, we provide recommendations for obtaining estimates for outcome and covariate ICCs, including the use of historical data or databases of published studies [23]. We encourage researchers to routinely report outcome and covariate ICCs in published trials to facilitate future sample size calculations. Finally, we showed how our calculator may be used to compare sample size and power for multiple designs. While this tutorial does not discuss power calculation under marginal models, the formulas presented here should be equally applicable to marginal models with continuous outcomes if the random effects structure matches the marginal model working correlation structure.

There are several areas for future work. First, our calculator currently supports testing for a single covariate only. Investigators may be interested in subpopulations defined by multiple covariates, necessitating a joint test of a global null across all interaction parameters. While such methods have been developed for two-level parallel CRTs [5], further work is required to apply them to longitudinal CRTs and CRXOs. Second, incomplete designs have been explored for ATE-focused CRTs [33] but, to our knowledge, have not yet addressed HTE analyses; power calculations in this setting would require a known matrix of all individual covariate values and, currently, is only possible via simulation. Third, including random treatment effects across clusters in our current model for power analysis requires future work, possibly through simulation-based calculations [34]. Additionally, decay correlation models have been developed for ATE-focused CRTs but have not been widely studied for HTE analyses. Finally, we note that while we focus on cross-sectional and closed-cohort sampling schemes here, many of the methods covered may also be extended to open-cohort designs and require future work [35].

## Ethics approval

Not applicable.

## Author contributions

MRB designed and coded the Shiny calculator, wrote the first draft of the manuscript, and revised subsequent drafts. FL directed the project. FL and MT assisted in drafting and revising the manuscript. PJH, MOH, GT, and RW provided detailed comments and revisions both to the manuscript and the Shiny calculator.

## Conflict of interest

None declared.

## Funding

Research in this article was supported by a Patient-Centered Outcomes Research Institute Award (PCORI®) Award ME-2020C3-21072. The statements presented in this article are solely the responsibility of the authors and do not represent the views of PCORI®, its Board of Governors or Methodology Committee.

## Data availability

No new data were generated or analyzed in support of this research. Source code for our CRT HTE Shiny Calculator can be found on Github: https://github.com/maryryan/CRT-HTE-calculator-app.

## Use of Artificial Intelligence (AI) tools

No AI tools were used during the course of this project.

# Supplementary material for "A tutorial on conducting sample size and power calculations for detecting treatment effect heterogeneity in cluster randomized trials with linear mixed models"

## Supplemental figures

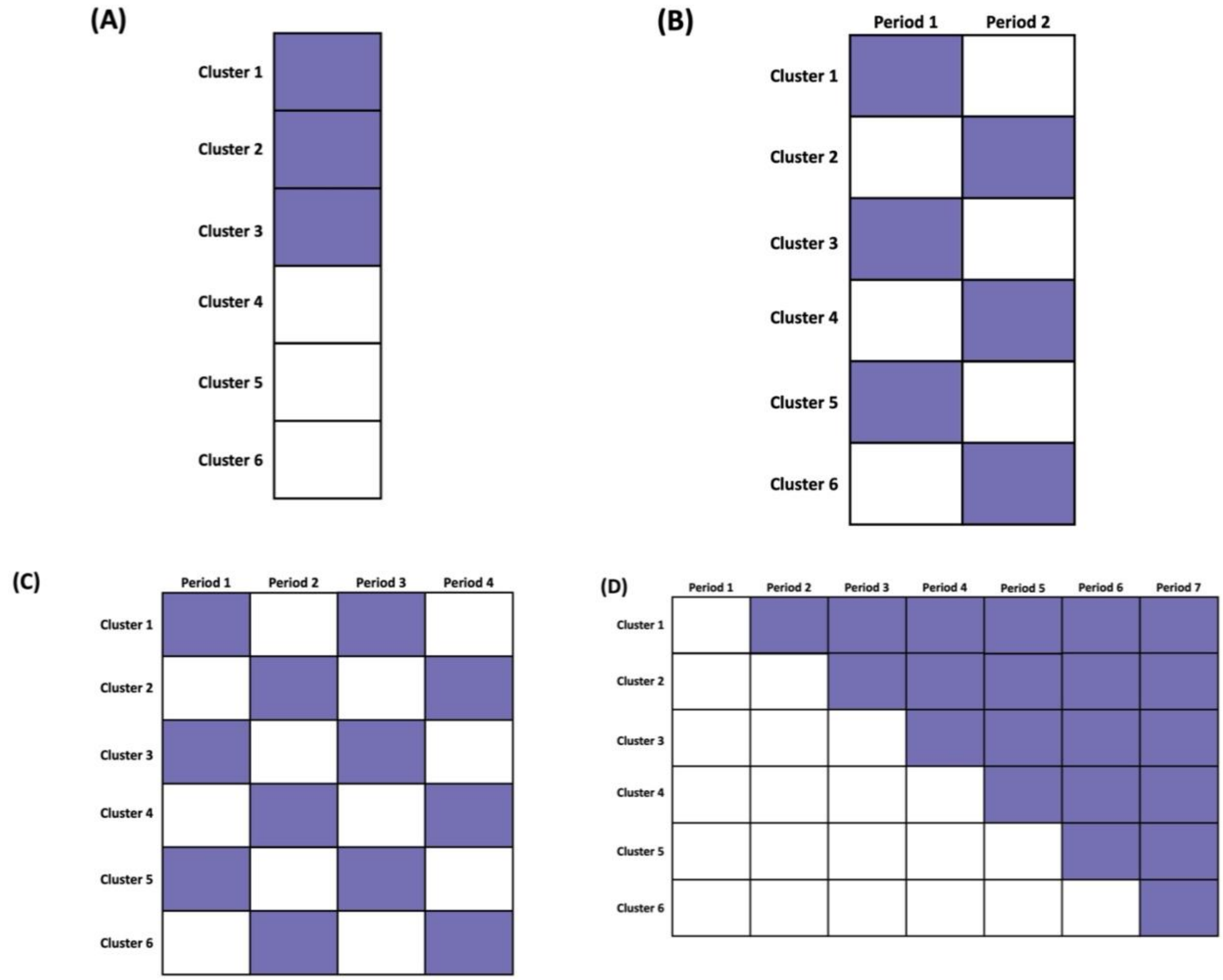

*Figure S1: Schematic examples of different types of cluster randomized trial (CRT) designs including: (A) two-level parallel, (B) cluster crossover, (C) multiple-period cluster crossover, and (D) stepped-wedge CRT.*

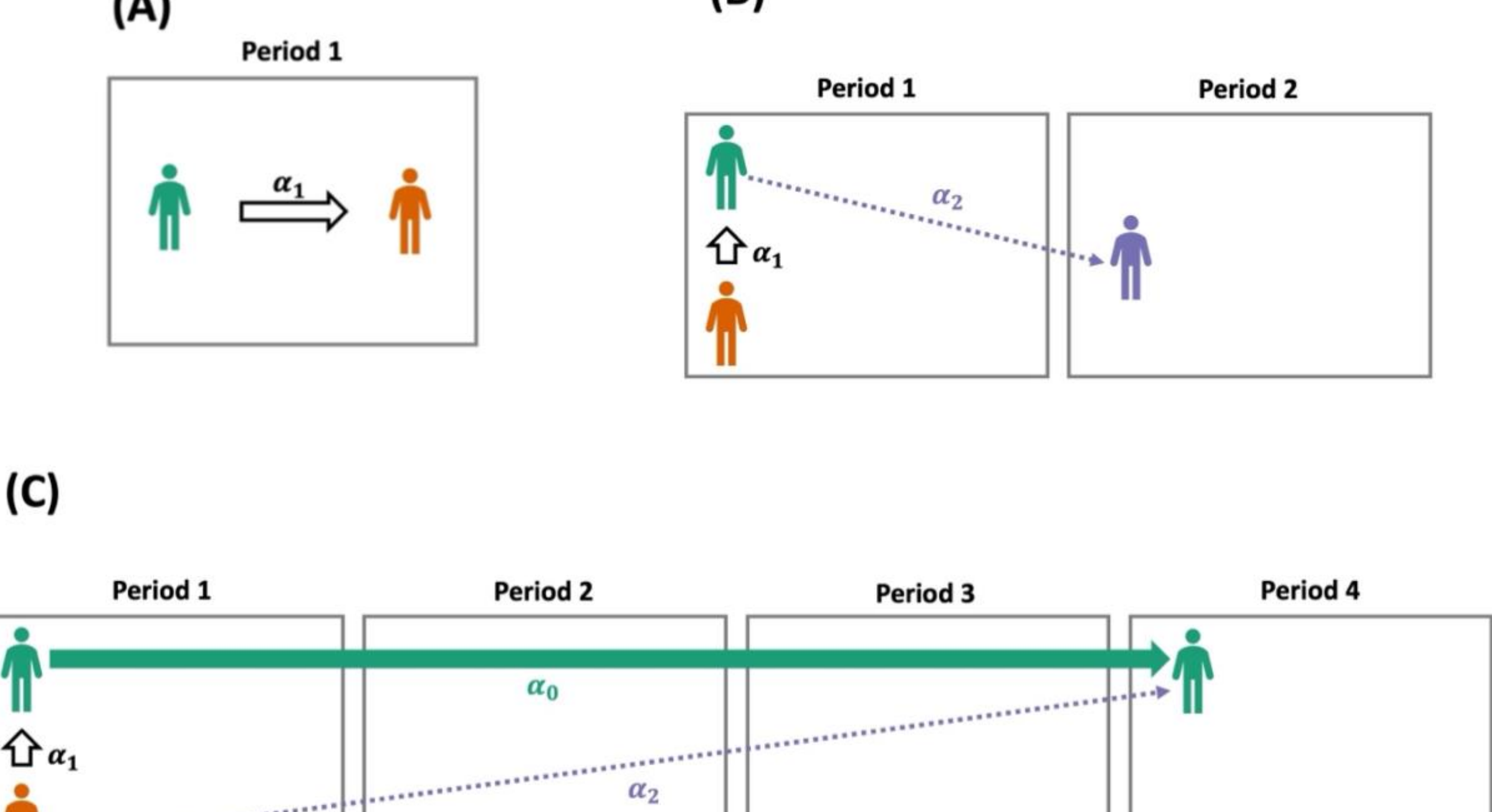

*Figure S2: Schematics of (A) an exchangeable correlation structure in a single cluster in a single time period, (B) a nested exchangeable correlation structure between cross-sectionally sampled individuals in a single cluster over two time periods, and (C) a block exchangeable correlation structure between longitudinally measured individuals in a single cluster over four time periods.*

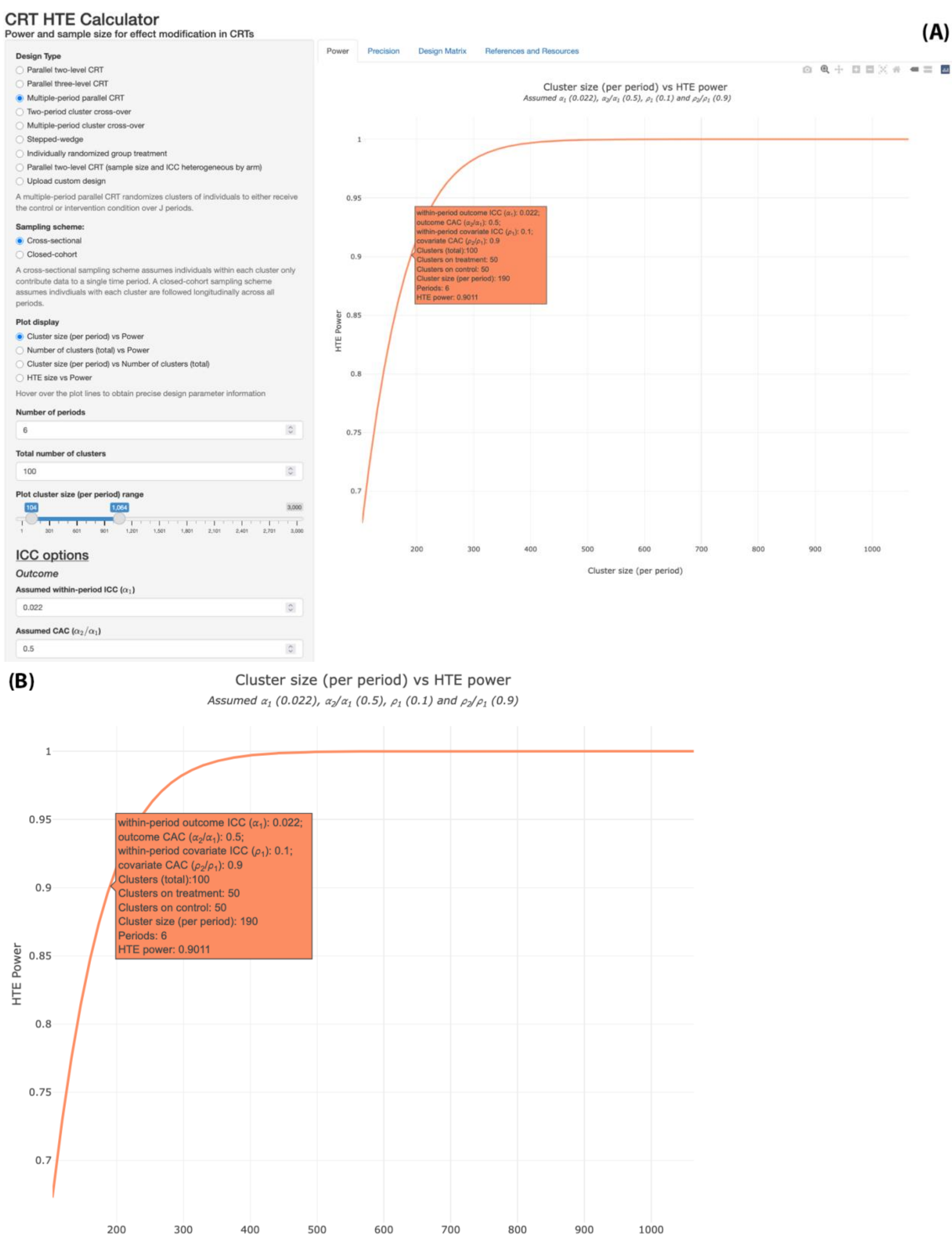

(A)
CRT HTE Calculator
Power and sample size for effect modification in CRTs
Design Type
Parallel two-level CRT
Parallel three-level CRT
Multiple-period parallel CRT
Two-period cluster cross-over
Multiple-period cluster cross-over
Stepped-wedge
Individually randomized group treatment
Parallel two-level CRT (sample size and ICC heterogeneous by arm)
Upload custom design
A multiple-period parallel CRT randomizes clusters of individuals to either receive the control or intervention condition over J periods.
Sampling scheme:
Cross-sectional
Closed-cohort
A cross-sectional sampling scheme assumes individuals within each cluster only contribute data to a single time period. A closed-cohort sampling scheme assumes indivdiuals with each cluster are followed longitudinally across all periods.
Plot display
Cluster size (per period) vs Power
Number of clusters (total) vs Power
Cluster size (per period) vs Number of clusters (total)
HTE size vs Power
Hover over the plot lines to obtain precise design parameter information
Number of periods
6
Total number of clusters
100
Plot cluster size (per period) range
104
1,064
3,000
1 301 601 901 1,201 1,501 1,801 2,101 2,401 2,701 3,000
ICC options
Outcome
Assumed within-period ICC ($\alpha_1$)
0.022
Assumed CAC ($\alpha_2/\alpha_1$)
0.5
Power
Precision
Design Matrix
References and Resources
Cluster size (per period) vs HTE power
Assumed $\alpha_1$ (0.022), $\alpha_2/\alpha_1$ (0.5), $\rho_1$ (0.1) and $\rho_2/\rho_1$ (0.9)
HTE Power
1
0.95
0.9
0.85
0.8
0.75
0.7
within-period outcome ICC ($\alpha_1$): 0.022;
outcome CAC ($\alpha_2/\alpha_1$): 0.5;
within-period covariate ICC ($\rho_1$): 0.1;
covariate CAC ($\rho_2/\rho_1$): 0.9
Clusters (total):100
Clusters on treatment: 50
Clusters on control: 50
Cluster size (per period): 190
Periods: 6
HTE power: 0.9011
200 300 400 500 600 700 800 900 1000
Cluster size (per period)
(B)
Cluster size (per period) vs HTE power
Assumed $\alpha_1$ (0.022), $\alpha_2/\alpha_1$ (0.5), $\rho_1$ (0.1) and $\rho_2/\rho_1$ (0.9)
HTE Power
1
0.95
0.9
0.85
0.8
0.75
0.7
within-period outcome ICC ($\alpha_1$): 0.022;
outcome CAC ($\alpha_2/\alpha_1$): 0.5;
within-period covariate ICC ($\rho_1$): 0.1;
covariate CAC ($\rho_2/\rho_1$): 0.9
Clusters (total):100
Clusters on treatment: 50
Clusters on control: 50
Cluster size (per period): 190
Periods: 6
HTE power: 0.9011
200 300 400 500 600 700 800 900 1000
Cluster size (per period)

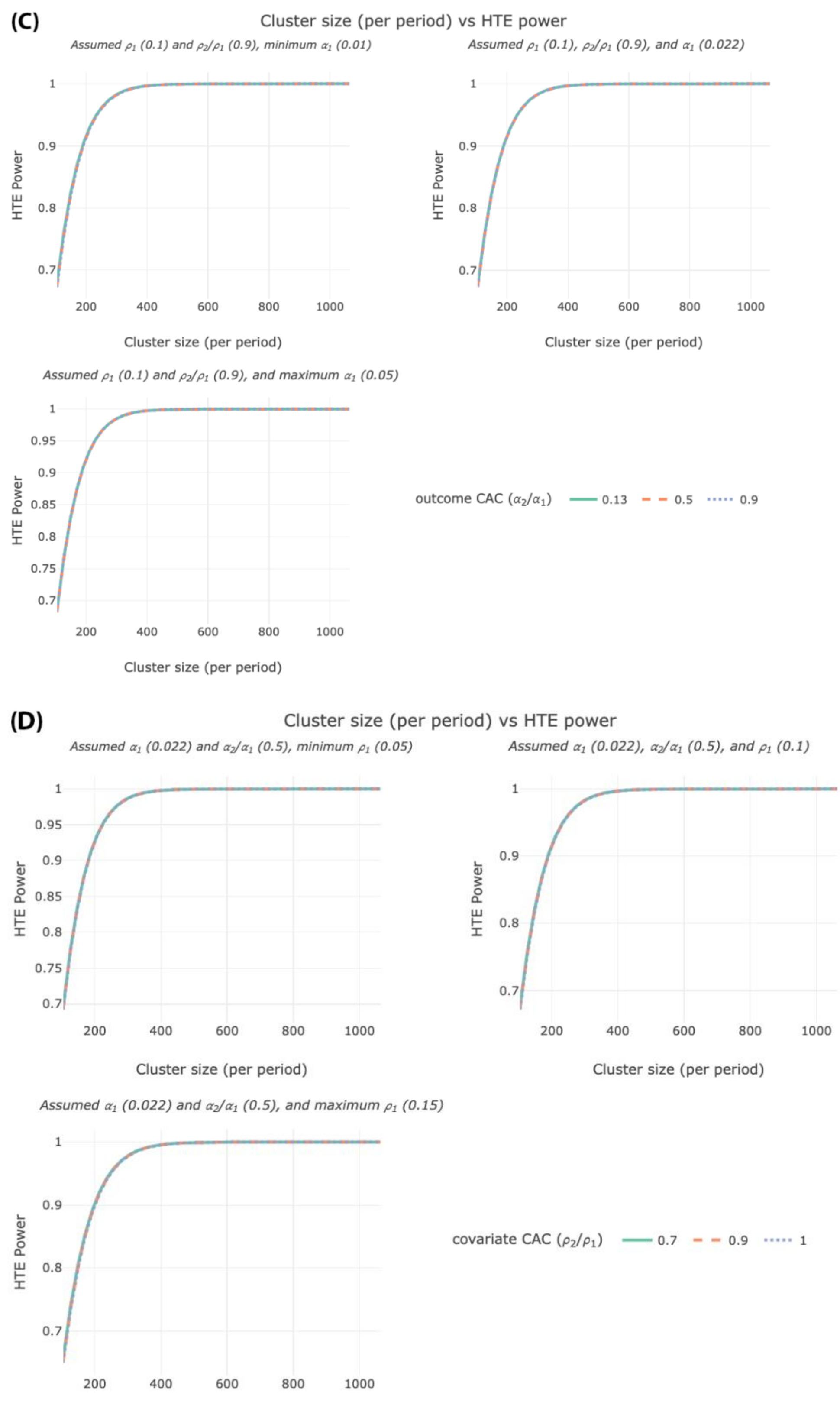

*Figure S3: Screenshot of the CRT HTE Shiny Calculator interface and power curves from calculator for a balanced six-period parallel cluster randomized trial (CRT) with a continuous outcome and a binary effect modify covariate, modeled after the Lumbar Imaging with Reporting of Epidemiology (LIRE) study. Scenario includes 6 time periods, 100 clusters (50 per arm), outcome intracluster correlation (ICC) of 0.022 (lower value 0.01 and higher value 0.05), outcome cluster autocorrelation coefficient (CAC) of 0.5 (lower value 0.13 and higher value 0.9), covariate ICC of 0.1 (lower value 0.05 and higher value 0.15), covariate CAC of 0.9 (lower value 0.7 and higher value 1), standardized heterogeneity of treatment effect (HTE) size of -0.05, effect modifier prevalence of 20%, and 0.05 significance level. Figure 1A shows the Figure 1B depicts curve assuming fixed within-period outcome ICC of 0.022, outcome CAC of 0.5, within-period covariate ICC of 0.1, and covariate CAC of 0.9. Figure 1C depicts multiple curves assuming within-period outcome ICC and CAC ranges of (0.01, 0.05) and (0.13, 0.9), respectively, a fixed within-period covariate ICC of 0.5, and a fixed covariate CAC of 0.9. Figure 1D depicts multiple curves assuming within-period covariate ICC and CAC ranges of (0.05, 0.15) and (0.7, 1), respectively, a fixed within-period outcome ICC of 0.022, and a fixed outcome CAC of 0.5.*

## Data example: exploration of a parallel CRT via the Umea Dementia and Exercise (UMDEX) study

The Umea Dementia and Exercise (UMDEX) study is a parallel cluster randomized trial (CRT) which evaluated the efficacy of a high-intensity functional exercise program to a seated control activity for older people with dementia in residential care facilities in Sweden [1]. Older adults living on the same floor, wing, or unit were randomized as a cluster to receive either treatment or control. The primary outcome was independence in activities of daily living (ADLs) as measured by the continuous motor domain of the Functional Independence Measure (FIM).

A further question of interest may be whether the intervention program effect on FIM differs by dementia type, which we can categorize as having an Alzheimer's disease (AD) or non-AD dementia diagnosis. To explore this question, we will use a parallel two-level CRT similar to UMDEX's original design. The parallel nature of the design makes an exchangeable correlation structure the most obvious choice for both the outcome and effect modifier. To estimate the number of clusters required, parameters we will need to specify include: cluster size ($m$), outcome intracluster correlation (ICC) ($\alpha_1$), covariate ICC ($\rho_1$), prevalence of the effect modifier, power threshold, and heterogeneity of treatment effect (HTE) size.

In the original UMDEX study, cluster sizes ranged from 3 to 8 participants each; in the design of our study, we will take this as a feasibility constraint and target cluster sizes near this range. As a conservative power estimate for the HTE, we will assume that cluster sizes do not vary by cluster. For choice of outcome ICC, we may use data from the original UMDEX as both will share the same outcome variable. UMDEX estimated its initial sample size under an assumed ICC of 0.02, while an outcome ICC of 0.04 was reported in study results; thus, we may explore an outcome range between 0 and 0.04. An ICC for the effect modifier, dementia type, was not reported, and reliable external data were not available to estimate it; therefore, we will assume a value of 0.2 with a wide range between 0 and 0.8 for illustration. To estimate the prevalence for our binary effect modifying variable, we may use UMDEX results which reported 36% of its participants had an AD diagnosis. Finally, a 5% type I error rate and standardized effect size of 0.7 are considered here for illustration.

Under these conditions, we find that 90% power is achieved with 35 clusters of 11 participants each ($n \times m = N = 385$), or 48 clusters of size 8 ($N = 384$). If the outcome ICC reaches its upper bound of 0.04, the sample size is minimally affected: 39 clusters of size 10 ($N = 390$) or 55 clusters of size 7 ($N = 385$) would be sufficient to achieve the same power.

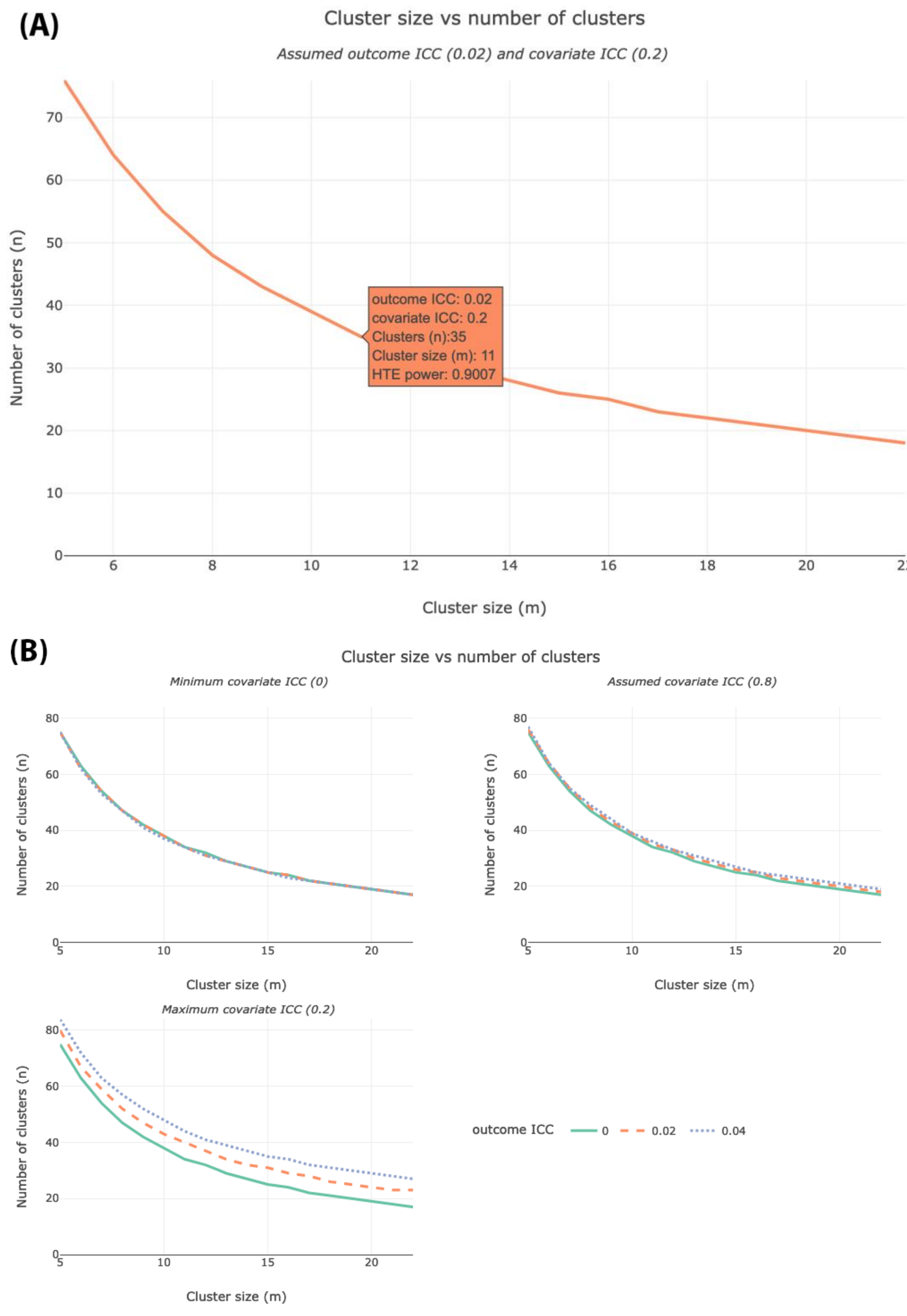

*Figure S4: Sample size curves from the CRT HTE Shiny Calculator for a two-level parallel CRT with a continuous outcome and a binary effect modifier, modeled after the Umea Dementia and Exercise (UMDEX) study. Scenario includes 90% power, standardized heterogeneity of treatment effect (HTE) size of 0.7, effect modifier prevalence of 36%, 1:1 intervention allocation, and 0.05 significance level. Figure S3A depicts*

*curve assuming fixed outcome intracluster correlation (ICC) of 0.02 and covariate ICC of 0.2. Figure S3B depicts multiple curves assuming an outcome ICC range of (0, 0.04) and covariate ICC range of (0, 0.8).*

An alternative design that may be under consideration is a parallel CRT with a baseline outcome measurement for both arms. This this case, two additional parameters would be required: a within-individual outcome ICC ($\alpha_0$), which we might assume to be moderate at 0.7, and the cluster autocorrelation coefficient (CAC), which we might assume to be relatively larger at 0.9. This assumes a closed-cohort sampling scheme over the baseline and active trial periods, and the design could be implemented in the calculator by uploading a 2x2 design matrix with a 1 in the lower right corner and zeroes elsewhere. In this case, the study would achieve 90% power with 32 clusters of size 6 ($N = 192$) or 18 clusters of size 11 ($N = 198$).

## Implications of HTE variance components

We now highlight the impact of HTE variance components on study power. In two-level CRTs, $\sigma^2_{\text{HTE}}$ decreases with smaller $\rho_1$ and larger $\sigma^2_x$, but has a parabolic relationship with $\alpha_1$. That is, the design effect for testing HTE (compared to the cluster average treatment effect [ATE]) has an upper limit with respect to $\alpha_1$, meaning that there is a ceiling to how much larger $\sigma^2_{\text{HTE}}$ can be compared to $\sigma^2_{\text{ATE}}$ for the same study. This suggests that by inflating the sample size of an individually randomized trial to detect the ATE in a CRT, any additional sample size inflation needed to sufficiently power the HTE may be minimal. We refer readers to Yang et al. (2020) [2] (specifically Figure 2) for further details and illustration.

In parallel multiple-period CRTs, longitudinal individual sampling can be less powerful for detecting an HTE than a cross-sectionally-sampled multiple-period CRT [3]. Yet, a cluster randomized crossover (CRXO) with multiple crossovers and longitudinal sampling would be more powerful for detecting an HTE than a similar CRXO with multiple crossovers and cross-sectional sampling. Further, larger covariate Intracluster correlations ICCs generally produce less power to detect an HTE in a multiple-period CRT, while increases in $\alpha_1$ or $\alpha_2$ will affect power differently depending on the values of other design parameters [3].

## Practical considerations

In designing studies with treatment effect heterogeneity in mind, there are several additional practical concerns that may need to be addressed, including how to obtain advanced estimates of ICC parameters, how to modify HTE sample size requirements for non-primary analyses, and consideration of non-constant cluster sizes and small numbers of clusters.

By definition, ICCs can range from 0 to 1, but in practice, commonly reported ICC values for outcomes in community-based CRTs rarely exceed 0.25 [4–6]. Unlike other study design parameters, there is often limited publicly reported information available to aid in estimating outcome and covariate ICCs. One strategy is to consult recently-published databases of outcome ICCs from completed CRTs [7], or to utilize data available from published trials or observational studies to estimate outcome and covariate ICCs in similar settings [8]. The latter strategy may be the most useful overall, as many studies may collect data on similar covariates or secondary outcome measures even if they evaluate different primary outcomes. In addition, investigators should consider that length of time period may impact the strength of within-individual or between-period ICC estimates, and should ensure that any historical study information they are using in the planning of their trial is comparable in this aspect. While ideally sample size calculations should match the planned analysis approach, outcome and covariate ICCs for binary variables should be estimated using a linear probability model, not a logistic model, to obtain estimates on the proportions scale as there is still no consensus on how to obtain between-period ICC estimates for binary proportions [9,10]. A detailed tutorial on how to obtain ICCs for sample size calculation in longitudinal clustered designs is provided in Ouyang et al. (2023) [8]. Although covariate ICCs are generally less published, a recent example is in Ouyang et al. (2024) [11] who presented empirical ICC estimates for age, sex, and race from the 2018 USA Medicare data to inform CRT design in Alzheimer's and related dementias. In addition, we note that the outcome ICC is dependent on the model under consideration: in the case of exploration of HTE, the outcome ICC will be one where the effect modifying covariate is adjusted for. This means that the adjusted outcome ICC will likely be smaller than the unadjusted outcome ICC. In the event that research teams only have access to unadjusted outcome ICC information at the study planning stage, they may perform sensitivity analyses where the unadjusted outcome ICC represents the top of the sensitivity range and smaller ICC values may be used to explore the impact on sample size and power. We also refer readers to Kasza et al. (2023) [12] for guidance on how to obtain ICC parameter estimates for more complex correlation structures using available information from studies assuming simpler structures.

Further, in many cases investigators may be primarily interested in the ATE but still want to verify *a priori* that their estimated sample size adequately powers their pre-specified HTE hypothesis. In this case, investigators can obtain the power of an HTE hypothesis via:

$$\text{power} = \Phi\left(\frac{|\Delta|}{\sqrt{\sigma^2_{\text{HTE}}/n}} - Z_{1-\alpha/2}\right),$$

where $\Phi(\cdot)$ is the cumulative standard normal distribution function and $\sigma^2_{\text{HTE}}$ is the design-specific estimated heterogeneous treatment effect variance. If HTE hypotheses are investigated as secondary hypotheses, power thresholds need not be as strict as those for primary hypotheses and the type I error level $\alpha$ may be set to a different value than the typical 0.05.

Many tools used for power estimation and analysis of CRTs rely on large-sample or asymptotic theory, which may not be accurate when the sample size is limited. This phenomenon is generally driven by a limited number of clusters, which is concerning as investigators often report difficulty in recruiting at this level [13,14]. To mitigate this issue in the design phase, investigators can use a *t*-distribution instead of a normal distribution for power calculations. For ATE analyses in CRTs, it has been shown that setting the degrees of freedom to the number of clusters minus two performs well in small sample sizes, mimicking the degrees of freedom for a cluster-level analysis. However, the optimal choice of degrees of freedom for CRTs has not been thoroughly studied [3,15–18].

It is also important to note the impact of selecting an appropriately flexible correlation structure in sample size estimation. It has been shown for studies focused on the ATE that not allowing for distinct between-period ICCs in multiple-period CRTs results in artificially small sample size predictions [19].

Finally, many sample size calculations assume a constant cluster size; this may not always be reasonable [20,21]. For CRTs evaluating the ATE, variation in cluster size will always reduce study power compared to CRTs with the same total sample size but constant cluster sizes [22]. For parallel CRTs focused on HTE analyses, however, the impact on power depends on covariate and outcome ICCs. For example, cluster size variation will increase power if the covariate ICC is smaller than the outcome ICC and will have no effect on power if the covariate and outcome ICCs are equal [23]. In scenarios where cluster size variation decreases power, the magnitude of power loss differs for ATE analyses versus HTE analyses as well as the type of effect modifier used [23–28]. Further, studies have shown that variable cluster sizes minimally affect HTE analyses involving an individual-level effect modifier, though the impact is more pronounced when the effect modifier is at the cluster level, due to the large covariate ICC [23,24]. As explicit methods to adjust for variable cluster size are currently available for only a limited number of designs, our online calculator only considers constant cluster sizes within arm, which is generally adequate when the HTE analysis is based on an individual-level effect modifier.

## Supplementary material references